\def\plotfiddle#1#2#3#4#5#6#7{\centering \leavevmode
\vbox to#2{\rule{0pt}{#2}}
\includegraphics{#1}}
\documentstyle[11pt,aspconf,epsf]{article}
\renewcommand{\thepage}{\arabic{page}}
\begin{document}

\title{Line Width and Spectral Index Distributions as Evidence for
Axisymmetry in the Broad Line Regions of Active Galaxies}

\author{C.M. \ Rudge and D.J. \ Raine}
\affil{Astronomy Group, University of Leicester, University Road,
Leicester, LE1 7RH, UK.}

\begin{abstract}
We propose that the scatter in line width versus luminosity of the BLR
of AGN arises from a dependence on the line of sight to an axially
symmetric BLR. Adopting a simple model for the line width as a
function of luminosity and angle, and convolving this with the
observed luminosity function, allows us to predict a line width
distribution consistent with the available data. Furthermore, we use
the relation between the equivalent width of a line and the luminosity
in the continuum to predict an observed correlation between line width
and equivalent width. The scatter on this correlation is again
provided by angular dependence. We also show a viewing angle
dependence can produce the X-ray spectral index distribution. The
results have applications as diagnostics of models of the broad line
region and in cosmology.
\end{abstract}
\section{Introduction}
The observed range of FWHM for the broad emission lines in AGN could
be driven by various mechanisms. An obvious possibility is the
luminosity of the central source. However plots of line width against
luminosity for observed samples show no clear correlation between
these parameters. It would be natural to expect higher luminosity
sources to have higher velocities in the emission regions and thus
broader lines, but powerful sources such as 3C273 can have narrower
lines than humbler Seyfert galaxies. In the context of `unified
models' it is natural to associate this parameter with viewing angle
to an axially-symmetric system. In the case of radio-loud sources
there is clear evidence of a relation between the radio core to lobe
dominance parameter, R, and FWHM (Wills \& Browne 1986). The lack of
spherical symmetry has been modelled in various ways including a simple
Kepler disc, an axisymmetric broad line cloud distribution, and a
spherical cloud distribution with non spherical illumination by the
central source. The `unified model' may involve a preferred axis of
illumination along the axis of symmetry, extending the observed
illumination cones of the ENLR into the BLR.  Thus, we develop a relation
for FWHM dependent on both viewing angle and luminosity. We shall
proceed initially in as model-independent a way as possible. Later we
shall return to the possibility that we can use the results to
discriminate between models.

Since we cannot measure the angle of inclination of the nuclear
regions, at least with any certainty, we resort to statistical
considerations. If the line width is a function of angle and luminosity
then averaging over angle and convolving with the luminosity function
will give a line width distribution which can in principle be compared
to observations. While the data are not sufficiently extensive to
provide a strong test there are several surveys available to validate
the general approach.

Our simplest description of the line width dependence contains three
parameters which the line width differences in a given source imply are
wavelength dependent. However, we can constrain the models further in
several ways. We know that the equivalent width of a line is related
to luminosity; this is the Baldwin relation. Since the slope of the EW
-- luminosity relation in a single time-variable nucleus is generally
different from the slope of the Baldwin relation this will show some
scatter. However, this is small compared to the scatter in the EW--FWHM
diagram. Ignoring the scatter in the Baldwin relation we show that the
part of the EW--FWHM plot that is populated by sources coincides with
that to be expected on the basis of the angle dependence of the FWHM
-- luminosity relation.

In the context of a unified model with an obscuring torus we would
also expect the X-ray continuum slope to be dependent upon viewing
angle. This is because the obscuring material preferentially absorbs
soft X-rays giving a harder spectrum in sources viewed edge-on. We
construct a similar angle-dependent functional form for the
dependence of X-ray slope on angle and luminosity and show that this
is consistent with the observed distribution of slopes.

We illustrate two applications. First as a diagnostic tool we show
that both the dual winds model (Cassidy \& Raine 1996) and the disc
wind model (Chiang \& Murray 1996) have an angular dependence of
line width consistent with the observed line width distribution function
but that a Keplerian disc does not. The extreme form of the model in
which line widths depend on angle only and not on luminosity is ruled
out. The line width dependence for the magnetic winds model as shown in
figure 6 of Emmering et al.\ (1992) is not consistent with this
model. We have not considered the luminosity dependence of line width
in these models but this might provide a more discriminating
diagnostic.

Finally we show the preliminary results, but not the details, of a
cosmological application to determine the deceleration parameter $q_{0}$
(in principle, given enough data).

\section{Angular Dependence}

We restrict attention to an axisymmetric BLR, so ignoring, for
example, warped discs. Line width ,$v$, taken to be a function of
viewing angle, $i$, and luminosity, $L$, can be written in general as
a double series of the form
\begin{equation} 
v(i,L_{44})=L_{44}^{\alpha }\sum_{n=0}^{\infty}\sum_{m=0}^{\infty
} a_{nm}L_{44}^{m}\cos n\theta  
\label{equn1}
\end{equation} 
where $i=\pi/2 - \theta$ is measured from the axis of symmetry,
$L_{44}$ is the luminosity in units of 10$^{44}$ erg s$^{-1}$
and $a_{nm}$ and $\alpha $ are parameters (independent of $L$ and
$\theta $). We use only the cosine series to impose symmetry about
the equatorial plane. Our model dependence enters only through

(i) the assumption that we can truncate this series, retaining only
the terms $a_{00}=a,$ and $a_{10}=b$. Thus equation (\ref{equn1})
simplifies to

\begin{equation}
v(i,L_{44})=(a+b\sin i)L_{44}^{\alpha },  
\label{vel}
\end{equation}
with $a$, $b$ and $\alpha$ constant for a given line.

(ii) We assume also a maximum value for inclination $i=i_{\ast}$. This
corresponds to the opening angle of the torus in the unified model. We
find that best fits are obtained for $i_{\ast}=60^{\circ}$ which,
while larger than many estimates of the opening angle, is consistent
with more recent observations and also the dual winds model (Cassidy
\& Raine 1996). For values of $i<i_{\ast}$ we assume systems are
randomly oriented.

\begin{figure}
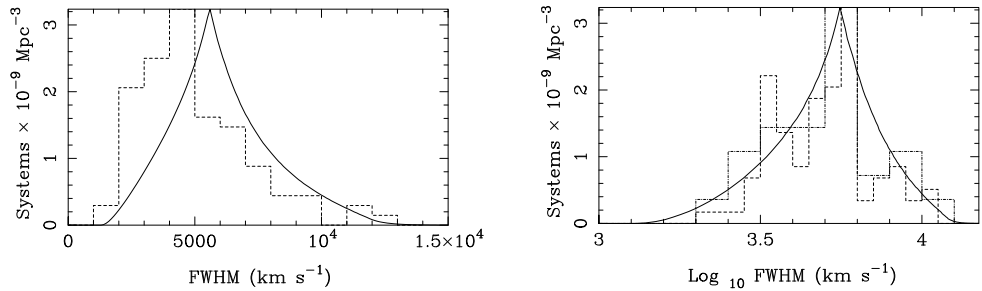

	\plotfiddle{rudge_fig1a.eps}{65pt}{-90}{25}{25}{-190}{100}
	\plotfiddle{rudge_fig1b.eps}{0pt}{-90}{25}{25}{10}{125}
	\caption{Model predictions for C\,{\sc iv} line width
	distribution overlaid
	with observed data by Wilkes (1987) (left) and
	Wills et al.\ (1993) (dashed)
	and Brotherton et al.\ (1994) (dot-dashed).}
	\label{c4lwd}
\end{figure}

We obtain a line width distribution by averaging over angle and
luminosity. Initially for modelling the line width distribution we use
a 2 power-law fit to the B-band luminosity function of Boyle et
al.\ (1988)
\begin{equation} 
\Phi (L_{44})\propto L_{44}^{-p} 
\left\{ 
\begin{array}{c} 
p=3.85\ \rm{ for }\ L_{44}>2 \\ 
p=1.27\ \rm{ for }\ L_{44}<2 
\end{array}
\right. . 
\label{lumfun} 
\end{equation}
However, for later work on the X-ray spectral index and cosmology, we
will use the X-ray luminosity function of Boyle et al.\
(1994). The line width distributions can be fitted equally well with
the X-ray luminosity function using different values of $a$, $b$ and
$\alpha$ which is consistent with predictions that $L_{x} \propto
L_{opt}^{\beta}$ for some value of $\beta$.  The number of systems
with line widths between $v$ and $v+$d$v$ is $N(v)$d$v$ and is given by
\[ 
N(v)=\int_{L_{44}}\Phi (L_{44})\sin i \frac{{\rm d} i}{{\rm d} v}
{\rm d}L_{44}. 
\]
This integral is evaluated numerically. Figure \ref{c4lwd} shows the
results for C{\sc iv}\,$\lambda 1549$ compared to observational
samples (Wilkes 1987, Wills et al.\ 1993 and Brotherton et al.\
1994). The Wilkes data is taken from the Parkes flat spectrum survey
which is biased towards face-on objects. Thus we expect the peak in
the distribution for this sample to lie to the low velocity side of
the model. The values of $a$ (the spherical component of velocity) and
$b$ (essentially the dipole component) are $a=2500$ km s$^{-1}$ and
$b=9500$ km s$^{-1}$ with $\alpha=0.15$. We have also obtained fits
for Mg{\sc ii}\,$\lambda 2800$ and H$\beta$ using data from the RIXOS
sample (Puchnarewicz et al.\ 1997). Because the data is biased it is
difficult to assess the agreement with the model statistically, but it
seems highly likely that a pure disc model, ($a=0$)  which would
predict a large number of narrow lined systems, and a spherical model
($b=0$), which would give a narrow peaked distribution, are ruled out.
Rudge and Raine (1998) showed that the angle dependence of line widths
in both the disc winds model of Cassidy \& Raine (1996) and the disc
wind model of Chiang and Murray (1996) can be fitted to the form
(\ref{vel}).

\begin{figure}
	\plotfiddle{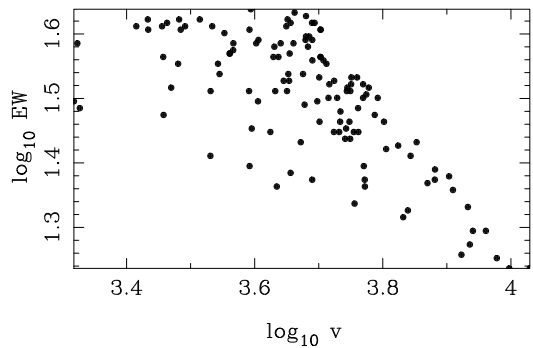}{125pt}{-90}{30}{30}{-110}{150}
	\caption{FWHM -- EW plot for C{\sc iv}.}
	\label{wew}
\end{figure}

The Baldwin relation for the line equivalent width, $EW$, is 
\[ EW\propto L^{\beta }\]
where for these lines $\beta_{{\rm C~IV}}=-0.17$ (Peterson, 1997), $\beta
_{{\rm H}\beta}=0.4$ and $\beta_{{\rm Mg~II}}=-0.37$ (Puchnarewicz, unpublished).
We have $v_{{\rm FWHM}}\propto L^{\alpha }$ with a large scatter from the
angular dependence. We therefore predict that 
\[
EW \propto v_{{\rm FWHM}}^{\beta /\alpha } 
\] 
with a large scatter. We construct a $\log EW-\log v_{{\rm FWHM}}$ plot for
systems drawn at random from a population given by the luminosity
function (equation \ref{lumfun}) and uniformly distributed in angle in
the range $0<i<i_{\ast}$. This is shown in figure \ref{wew} (see
figure 7 of Wills et al.\ (1993) for comparison with observations). We would
expect the time dependence of the Baldwin relation (and possibly a
gradual rather than a sudden cut-off near $i\sim i_{\ast}$) to smooth out
the upper edge of the predicted distribution. The predicted slope of the
upper boundary is $\beta /\alpha =-1.1$ compared with $-1\pm 0.25$
from inspection of Wills et al.\ (1993). Similar plots can be obtained
for Mg{\sc ii} and H$\beta $ with slopes 1.1 and 1.3 respectively,
compared with $ 1.6\pm 0.4$ and $0.6\pm 0.4$ (Puchnarewicz,
unpublished).

\section{X-ray Slopes}

We now apply a similar idea to the slope, $\alpha_{x}$, of the X-ray
spectra. The analytical form we choose is motivated by the following
argument. Assume that the hard X-ray luminosity $L_{h}$ at $\nu _{h}$
is independent of angle and the soft component $L_{s}$ centered on $\nu
_{s}$ has the form (\ref{vel}). Then
\begin{equation}
\alpha _{x}=\frac{\log (L_{s}/L_{h})}{\log (\nu _{h}/\nu _{s})}\propto \log
[(c + d \sin i)L_{44}^{\beta}]
\label{alphax}
\end{equation}
(with $d<0$ to accommodate softer spectra in face-on
systems). As above we can derive the corresponding $\alpha_{x}$
distribution by integrating over the luminosity distribution and
angle. In figure 4 we show the theoretical curve for $c=4.0,$ $d=-3.0$
and $\beta=-0.8$, overlaid on the spectral index distribution for the
RIXOS (Puchnarewicz et al.\ 1997), EMSS (Ciliegi \& Maccacaro 1996)
and Wandel \& Boller (1998) objects. It can be seen that the
Wandel \& Boller objects are not fitted by the model curve. This is
due to the sample having a large number of NLS1's which have in
general a softer X-ray spectrum. Combining equations
(\ref{vel}) and (\ref{alphax}) we obtain an expression for line width
dependent upon spectral index and luminosity. This obviously predicts
that face-on sources with narrower lines also have softer
X-ray spectra, as is shown in Wandel \& Boller (1998). However
their spherically symmetric BLR, which models this relation, does not
reproduce the observed line width distribution (Rudge \& Raine in
prep.).

\begin{figure}
	\plotfiddle{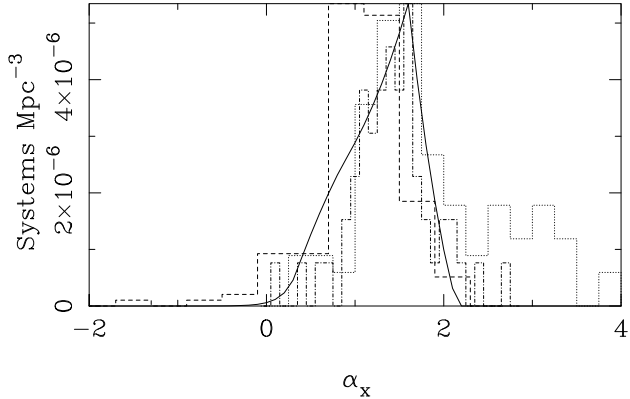}{130pt}{-90}{35}{35}{-150}{175}
	\caption{Model predictions for spectral index distribution
	overlaid with RIXOS (dashed), EMSS (dot-dashed) and Wandel
	\& Boller (dotted) data.}
	\label{ax}
\end{figure}

\section{Application to Cosmology}
Since the analysis enables us to account for the scatter in the
FWHM-luminosity relation for AGN it should be possible in principle to
use the model as a cosmological test. To indicate the results that
might be expected we show the difference in evolution of $N(v)$ from
$z<0.5$ to $2.5<z<3.0$ for $q_{0} =0.5$ and $q_{0} =0$, ($H_{0}=50$ km
s$^{-1}$Mpc$^{-1}$). Of course, for practical application the test
requires a large sample of line widths at high redshift.
\begin{figure}
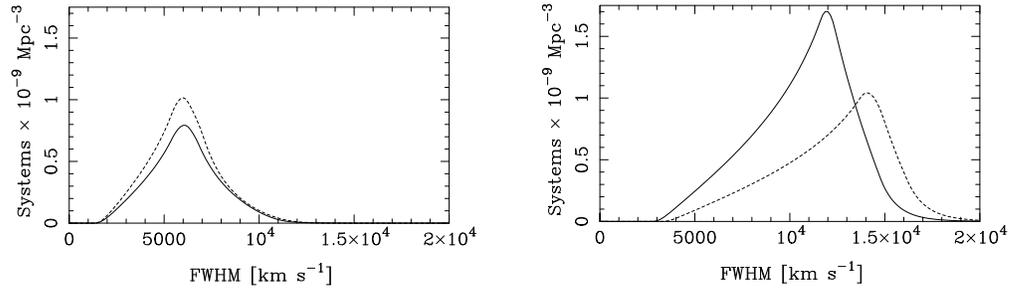

	\plotfiddle{rudge_fig4a.eps}{65pt}{-90}{25}{25}{-190}{100}
	\plotfiddle{rudge_fig4b.eps}{0pt}{-90}{25}{25}{10}{125}
	\caption{Line width distribution for $0<z<0.5$ (left) and
	$2.5<z<3$ (right). Solid line is for $q_{0}=0.5$ and dashed
	line is for $q_{0}=0$. Model parameters $a$, $b$ and
	$\alpha$ are the same in all cases.} 
	\label{cosm}
\end{figure}

\acknowledgements
CMR acknowledges the support of PPARC in the form of a Research
Studentship. The authors wish to thank Gordon Stewart and Adam Blair
for assistance with modeling the luminosity function and also Liz
Puchnarewicz and Amri Wandel for providing data.

\begin{question}{Mark Bottorff}
For the magnetic winds model (Emmering et al.\ 1992) the figure shown
in their paper is a very specific case. Changing the input parameters,
particularly the initial angle of the clouds relative to the accretion
disc surface, can produce a $v$--$i$ relation in the opposite sense to
that shown i.e. consistent with this model.
\end{question}
\begin{answer}{Christopher Rudge}
A deeper investigation of this model could produce more detailed
information on the exact nature of the $v$--$i$ relation. While
changing the initial conditions of the clouds will obviously affect
the $v$--$i$ relation it is unclear how much variation is possible
whilst still producing reasonable emission line
profiles. Consideration should also be given to the ability of this
model to predict the EW--FWHM relation in the case of $v$ decreasing
with increasing $i$.
\end{answer}
\begin{question}{Michael Corbin}
The evidence for viewing angle effect based on the R parameter can
only be applied to radio loud objects. The line width distributions
for radio loud and radio quiet objects are significantly different in
my sample (Corbin 1997), even with the same magnitude limit,
indicating an intrinsic difference between radio types.
\end{question}
\begin{answer}{Christopher Rudge}
The FWHM--R relation is perhaps one of the better pieces of evidence
for a viewing angle effect but not the only piece of evidence
i.e. relation to spectral index. The observed sample given in Corbin
(1997) does show a small difference between the line width
distributions for the radio loud and radio quiet objects. However even
in the magnitude limited case, the luminosities of the radio loud
objects are slightly higher than for the radio quiets. This model
predicts that the distribution would therefore peak at a higher FWHM
for the radio loud objects.
\end{answer}
\begin{question}{Neil Brandt}
Were the spectral index model and the $\alpha_{X}$--$v$ relation
developed specifically to model the RIXOS objects or to investigate ,
more generally,  the Boroson and Green eigenvector? Note also that the
intrinsic absorption seen in the RIXOS sample was not observed in our
sample of NLS1's.
\end{question}
\begin{answer}{Christopher Rudge}
While the spectral index model was initially suggested by the study of
the RIXOS objects by Puchnarewicz et al.\ , we have also been able to
fit other samples with this model. Future work is expected to include
a study of the various parameters related to the Boroson and Green
eigenvector. There is some evidence that the NLS1's are predominantly
face on Seyfert 1's. If this is the case then our model would
correctly predict their narrow lines and the lack of observed dust
absorption.
\end{answer}

\end{document}